\RequirePackage[table]{xcolor}
\documentclass{PoS}

\usepackage{latexsym,amssymb,amsmath,amsbsy,multirow}

\usepackage{stmaryrd,xcolor}
\usepackage{amsfonts,float,ulem}
\usepackage[pdftex]{graphicx}
\usepackage{bbm}
\usepackage{color}

\newcommand{\be}{\begin{equation}}
\newcommand{\ee}{\end{equation}}
\newcommand{\ba}{\begin{eqnarray}}
\newcommand{\ea}{\end{eqnarray}}
\newcommand{\bs}{\begin{subequations}}
\newcommand{\es}{\end{subequations}}

\graphicspath{ {figures/} }

\allowdisplaybreaks

\title{Machine Learning for Prediction of Unitarity and Bounded from Below Constraints}

\ShortTitle{Machine Learning for Prediction of Unitarity and Bounded from Below Constraints}

\author{\speaker{D.~Jur\v{c}iukonis} \\
		$^{(1)} \! $		
		Vilnius University,
		Institute of Theoretical Physics and Astronomy \\     		
        E-mail: \email{darius.jurciukonis@tfai.vu.lt}}

\abstract{The machine learning (ML) techniques to predict unitarity (UNI) and bounded from below (BFB) constraints in multi-scalar models is employed. The effectiveness of this approach is demonstrated by applying it to the two and three Higgs doublet models, as well as the left-right model. By employing suitable neural network architectures, learning algorithms, and carefully curated training datasets, a significantly high level of predictivity is achieved. Machine learning offers a distinct advantage by enabling faster calculations compared to alternative numerical methods, such as scalar potential minimization. This research investigates the feasibility of utilizing machine learning techniques as an alternative for predicting these constraints, offering potential improvements over traditional numerical calculations.
}

\FullConference{The European Physical Society Conference on High Energy Physics (EPS-HEP2023)\\
 21-25 August 2023\\
Hamburg, Germany\\}
		
\begin{document}

\section{Details of the computations}

Theoretical constraints such as unitarity and bounded from below conditions play important roles in model building within the realm of high energy physics, ensuring both the physical consistency and stability of the theoretical models. 
In high energy physics, UNI constrains the behavior of scattering amplitudes, ensuring that they remain finite and well-behaved. 
The BFB conditions is essential for the stability of the vacuum. 
In the context of scalar field theories, the BFB condition helps in ensuring that the scalar potential is phenomenologically consistent and that there is no direction in field space along which the potential tends to minus infinity.

The unitarity conditions can be computed analytically. Typically, the computation necessitates determining the eigenvalues of the scattering matrices. These computations are precise and fast.
The analytical computation of BFB conditions is feasible only for simple models. For complex models, minimization of the scalar potential is required. The computations involved in minimization are often slow and imprecise.

In this paper, three models with precise analytical procedures for BFB computations were investigated to verify the reliability of machine learning. 
Necessary and sufficient conditions for the scalar potential of the general two Higgs doublet model (G2HDM) to be BFB
were derived by several groups in ref.~\cite{Maniatis:2006fs,Ivanov:2015nea} through computation of simple and precise algorithm~\cite{Jurciukonis:2018skr}. 
The aligned three Higgs doublet model (A3HDM) has a quite large number of free parameters in the quartic scalar potential, but due to additional constraints, the BFB conditions for the model can be expressed analytically~\cite{Ferreira:2017tvy,Jurciukonis:2021wny}, while the CP-conserved left-right (LRM) model requires a quite complicated algorithm for the estimation of BFB conditions~\cite{Kannike:2021fth,Fontes:2022sjp}.

Typically, the parameters of quartic potential, lambdas, are generated within certain limits and later they are sampled according to UNI or BFB conditions~\footnote{Randomly generated sets of parameters are referred to as raw data, whereas samples that satisfy constraints are referred to as true samples.}.
Depending on the model, there may be a very small percentage of samples that satisfy UNI, BFB, or both UNI and BFB conditions, as shown in Table~\ref{table_percentage}. 
\begin{table}[ht]
  \begin{center}
    \begin{tabular}{|c|c|c|c|c|}
    \hline
     Model ($\lambda$'s) & \hspace*{3.mm} UNI \hspace*{3.mm} & \hspace*{3.mm} BFB-I \hspace*{3.mm} & \hspace*{3.mm} BFB-II \hspace*{3.mm} &
     UNI+BFB \\ \hline \hline
     G2HDM (10) & 1.02 & 23.3 & 3.25 & 0.033 \\
     LRM (13) & 0.41 & 18.4 & 0.7 & 0.0028 \\
     A3HDM (14) & 0.18 & 1.73 & 0.05 & 0.00009 \\     
         \hline
    \end{tabular}
  \end{center}
  \vspace*{-4mm} 
\caption{Percentage of true samples for models when lambdas are generated randomly. The number of model parameters to be generated is indicated in the brackets in the first column. The "BFB-I" column displays the percentage of true samples from raw data, whereas the "BFB-II" column indicates the percentage from datasets refined by UNI conditions. 
    \label{table_percentage}}
\end{table}

There are two computational strategies:
(1) Train networks to predict BFB constraints using data that satisfy UNI constraints, or
(2) Train networks to predict both UNI and BFB constraints simultaneously.
In this paper, we concentrate on the latter strategy.

For predictions of UNI and BFB constraints, we use four linear networks, each having eight layers. 
The initial network, net-1, consists of 128 neurons and is trained using raw data supplemented with an appropriate number of true samples. 
This network, net-1, is subsequently used to prepare the training data for net-2, which has 256 neurons. 
Both net-3 and net-4 are trained using the same data as net-2, but they employ larger matrices, with 512 and 1024 neurons respectively. 
These networks, net-3 and net-4, are then utilized to filter the predictions made by net-2.

For this research, a desktop computer was used.
For network training, we utilized a GPU (NVIDIA GeForce RTX 4090), whereas other computations were performed using a CPU (Intel® Core™ i9-13900K). 
Given the prevalent use of Wolfram Mathematica in the high energy physics (HEP) community, we chose to conduct our machine learning and other computations within Mathematica to ensure seamless integration with other HEP packages.
For training, we use $10^6-10^7$ samples with a batch size of $2^{14} = 16\,384$ and perform 500 training rounds for each network.

\section{Results}

\begin{figure}[H]
\begin{center}
\includegraphics[width=1.0\textwidth]{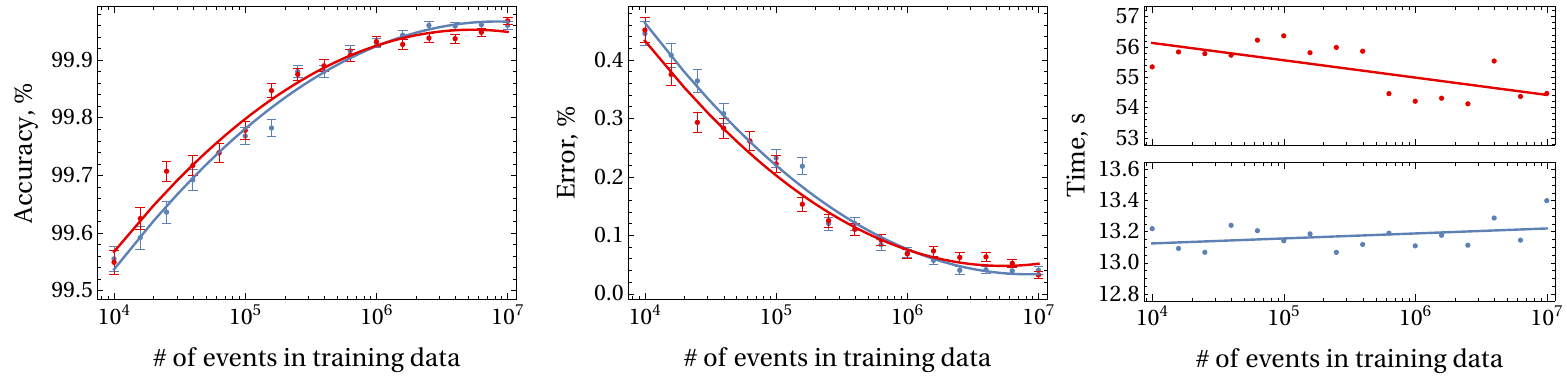}
\end{center}
\vspace{-4mm}
\caption{Classifier measurements (accuracy, error and prediction time) as functions from the size of raw training data. Blue points represent net-1, while red points correspond to measurements from net-4.}
\label{picture1}
\end{figure}

To predict across the widest possible parameter space, it is necessary to use a large training dataset. 
This is because the percentage of true samples in the raw data is very small.
As is typical, increasing the number of samples enhances the prediction accuracy as depicted in Figure~\ref{picture1}.
However, increasing the number of neurons in the networks can significantly increase the prediction time, but the accuracy does not improve, so an optimal network architecture must be chosen.
%

Assuming that the considered models do not have an analytical procedure for BFB calculation, we can evaluate them by minimizing the potential, for comparison with computations using neural networks. 
The prediction of UNI and BFB conditions using neural networks can be accelerated up to 70 times in the case of G2HDM, up to 100 times in the case of LRM, and up to 10 times in the case of A3HDM, compared to calculations when the model potential is minimized. 
The predictions from the networks can be recalculated using the minimization procedure, ensuring that computation times remain manageable.


When classifying data from raw data, even the simplest net-1 network exhibits very high accuracy, over 99.95\%, because raw data mainly consists of false samples. 
It is therefore interesting to compare which portion of the true samples classified by the neural network is correct. 
Consequently, we recalculate these true samples, using the analytical UNI and BFB conditions. 
We observe that in this case, the accuracy of net-1 is less than 50\%, but with more complex networks or their combination, the accuracy can be significantly improved, as shown in Table~\ref{table_percentage2}.
\begin{table}[ht]
  \begin{center}
    \begin{tabular}{|c|c|c|c|c|}
    \hline
     Model & \hspace*{3.mm} net-1 \hspace*{3.mm} & \hspace*{3.mm} net-2 \hspace*{3.mm} & \hspace*{3.mm} net-3,4 \hspace*{3.mm} &
     net-2 -- net-4 \\ \hline \hline
     G2HDM & 52-55 & 96-97 & 97-98 & $> 99$ \\
     LRM & 42-47 & 91-92 & 91-93 & $> 98$ \\
     A3HDM & 13-15 & 90-92 & 90-91 & $> 97$ \\
         \hline
    \end{tabular}
  \end{center}
 \vspace*{-4mm} 
\caption{Percentage of true samples within predicted results (from raw data), verified using analytical UNI+BFB conditions.
    \label{table_percentage2}}
\end{table}

\begin{figure}[H]
\begin{center}
\includegraphics[width=1.0\textwidth]{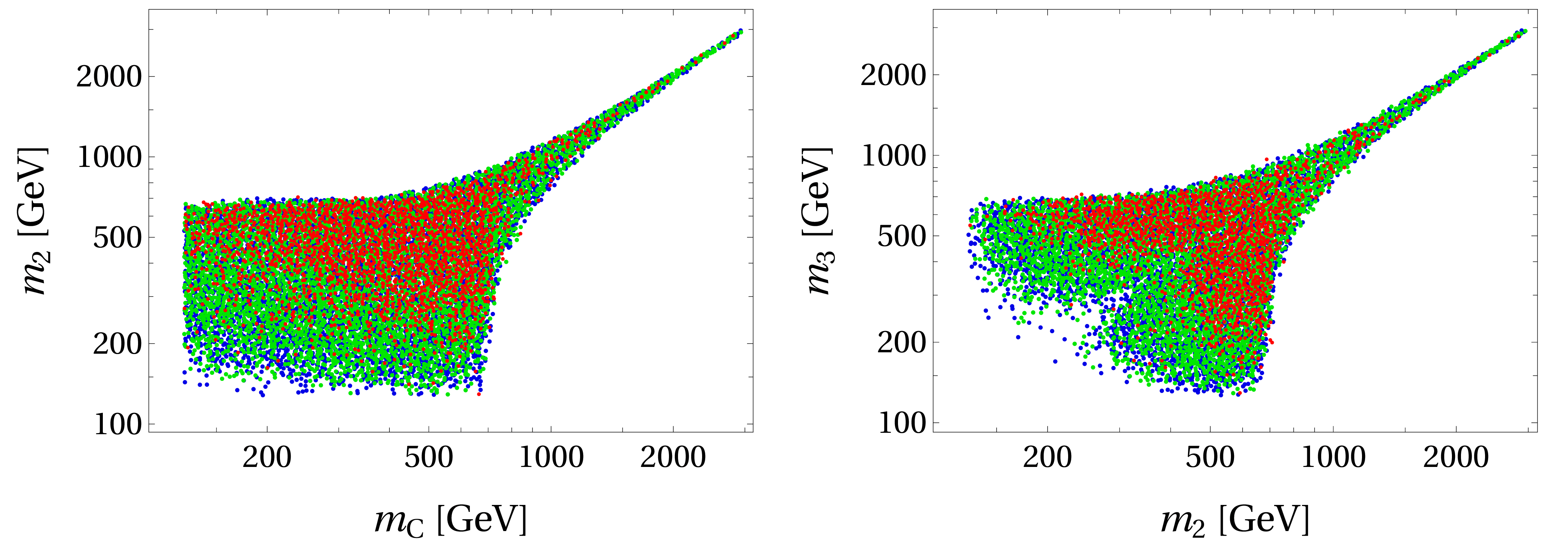}
\end{center}
\vspace{-6mm}
\caption{Scatter plots of Higgs masses for the G2HDM. The blue points represent correct computations based on analytical UNI+BFB conditions. The green points depict neural network predictions using the combined networks from net-2 to net-4, while the red points indicate false samples provided by net-2.}
\label{picture2}
\end{figure}

Since neural networks effectively classify UNI+BFB conditions, true samples can be used for further calculations, such as evaluations of various physical quantities. 
In Figure~\ref{picture2}, we see that the neural network predictions agree very well with the correct results. 
Additionally, false samples remain within the correlation ranges of the physical quantities avoiding anomalous values.
%


\paragraph{Concluding remarks:}
This analysis demonstrates that machine learning technique can effectively predict UNI and BFB constraints in multi-scalar models.
Simple linear networks can achieve high prediction accuracy, though they require sizeable training data samples.
Machine learning technique can significantly reduce computing time in comparison to the global minimization technique.


\end{document}